\documentclass[twocolumn, prl,showpacs,superscriptaddress]{revtex4-1} 
\usepackage{graphicx}
\usepackage{amsmath}
\usepackage{amssymb}
\usepackage{epstopdf}
\usepackage{dcolumn}
\usepackage{bm}

\newcommand*{\etal}{\emph{et al.}~}

\newcommand*{\tow}{$\lambda_\textrm{zero}=768.9712(7)_\textrm{stat}(8)_\textrm{sys}$ nm}

\begin{document}

\title{Measurement of a magic-zero wavelength}

\affiliation{Department of Physics, University of Arizona, Tucson, AZ 85721}
\affiliation{College of Optical Sciences, University of Arizona, Tucson, AZ 85721}
\author{William F. Holmgren}
\affiliation{Department of Physics, University of Arizona, Tucson, AZ 85721}
\author{Raisa Trubko}
\affiliation{College of Optical Sciences, University of Arizona, Tucson, AZ 85721}
\author{Ivan Hromada}
\affiliation{Department of Physics, University of Arizona, Tucson, AZ 85721}
\author{Alexander D. Cronin}
\affiliation{Department of Physics, University of Arizona, Tucson, AZ 85721}
\affiliation{College of Optical Sciences, University of Arizona, Tucson, AZ 85721}
\email{cronin@physics.arizona.edu}
\homepage{http://www.atomwave.org}

\date{\today}

\begin{abstract}
Light at a magic-zero wavelength causes zero energy shift for an atom.  We measured the longest magic-zero wavelength for ground state potassium atoms to be $\lambda_\textrm{zero}=768.971(1)$ nm, and we show how this provides an improved experimental benchmark for atomic structure calculations.  This $\lambda_\textrm{zero}$ measurement determines the ratio of the potassium atom D1 and D2 line strengths with record precision.  It also demonstrates a new application for atom interferometry, and we discuss how decoherence will fundamentally limit future measurements of magic-zero wavelengths.
\end{abstract}

\pacs{03.75.Dg, 32.10.Dk, 33.15.Kr}

\maketitle

The light-induced energy shift of an atom depends on the light wavelength, and there exist \emph{magic-zero wavelengths} for which the energy shift vanishes \cite{LeB07,Aro11}. A magic-zero wavelength ($\lambda_\textrm{zero}$) is found between atomic resonances, where the light is red-detuned from one resonance and blue-detuned from another. Opposing contributions from these resonances produce a root in the energy shift spectrum at $\lambda_\textrm{zero}$. In this Letter we report a  measurement of a magic-zero wavelength made with an atom interferometer.

LeBlanc and Thywissen \cite{LeB07} referred to $\lambda_\textrm{zero}$ as \emph{tune-out wavelengths} and discussed their utility for multi-species atom traps. Since then, various $\lambda_\textrm{zero}$ have been used in experiments to study entropy exchange \cite{Cat09}, quantum information processing \cite{Dal08}, and diffraction of matter waves from an ultracold atom crystal \cite{Gad12}.  However, the light used in experiments \cite{Cat09,Dal08,Gad12} to minimize energy shifts can be hundreds of picometers different than the $\lambda_\textrm{zero}$ values calculated in \cite{LeB07,Aro11} due to impure optical polarization. LeBlanc and Thywissen predicted a $\lambda_\textrm{zero}$ for each alkali atom with 10 pm precision based on the wavelengths of their principal (D1 and D2) transitions. More recently, Arora \etal \cite{Aro11} predicted magic-zero wavelengths using state-of-the-art atomic theory calculations of dipole matrix elements for several transitions in each atom, including core electron excitations.  For the $\lambda_\textrm{zero}$ we measured, Arora \etal stated a theoretical uncertainty of 3 pm. In comparison, our measurement has an uncertainty of 1 pm.  Because calculations of dipole matrix elements similar to those used in \cite{Aro11} are needed to calculate static polarizabilities, state lifetimes, line strengths, van der Waals potentials, and magic wavelengths \cite{Mit10,Der11,Lud08}, we are motivated to explore how measurements of magic-zero wavelengths can serve as new benchmark tests of atomic structure calculations.

In this Letter we present a measurement of the magic-zero wavelength for potassium between the 770 nm (D1) and 767 nm (D2) transitions. Our measurement of \tow~is a novel test of atomic structure calculations and provides the most precise determination yet of the ratio of the D1 and D2 line strengths $S_1$ and $S_2$. We find the ratio
\begin{eqnarray}
\label{Reqn}
R=\frac{S_2}{S_1}=\frac{|\langle 4s || D || 4p_{3/2}\rangle|^2}{|\langle 4s || D || 4p_{1/2}\rangle|^2}=2.0001(28).
\end{eqnarray}
The ratio of degeneracies for the excited states would make $R=2$, however, relativistic corrections slightly reduce the predicted ratio to $R=1.9987$ \cite{Saf12per}.  Our measurement is consistent with the prediction in \cite{Aro11}, and our measurement uncertainty is three times less than the theoretical uncertainty quoted in \cite{Aro11}.

Most measurements of static and dynamic polarizabilities \cite{Ami03, Dei08, Gou05, Mor93, Hac07} are limited by uncertainty in the electric field strength and uncertainty in the time an atom interacts with the field. However, our measurement of the wavelength at which the polarizability is zero is not subject to uncertainty from these factors. Instead, we will discuss systematic errors in $\lambda_\textrm{zero}$ measurements caused by laser spectra, and statistical limitations caused by contrast loss and small (mrad/pm) phase shifts near $\lambda_\textrm{zero}$.

The longest magic-zero wavelengths for alkali atoms are determined mostly by the transition energies $\hbar\omega_1$ and $\hbar\omega_2$ and the ratio $R$ of the line strengths. We use the sum-over-states approach to describe the dynamic polarizability $\alpha(\omega)$ near these two transitions by
\begin{eqnarray}
\label{dynpol}
\alpha(\omega) = & \frac{1}{3\hbar} S_1 \left( \frac{\omega_1}{  \omega_{1}^2 - \omega^2} + R\frac{\omega_{2}} { \omega_{2}^2 - \omega^2} \right) + A
\end{eqnarray}
where $A$ accounts for contributions from core excitations, higher energy valence transitions, and core-valence coupling \cite{Saf06, Mit10}. At the longest magic-zero wavelength of potassium, $A$ is 0.02\% of the nearly equal and opposite contributions from the principal transitions to the polarizability and $A$ changes $\lambda_\textrm{zero}$ by 0.15(1) pm \cite{Saf12per}. Therefore, the uncertainty in this magic-zero wavelength calculation is nearly entirely determined by uncertainty in the ratio of the line strengths, $R$.

The line strengths $S_1$ and $S_2$, and thus $R$, can also be determined from state lifetime measurements. To our knowledge, the most precise \emph{independent} measurements of the $4p_{1/2}$ and $4p_{3/2}$ state lifetimes were performed by Volz \etal using beam-gas-laser spectroscopy \cite{Vol96}. They reported lifetime uncertainties of 0.25\% and a similar uncertainty for $R$ (which leads to a 3 pm uncertainty in $\lambda_\textrm{zero}$). In comparison, our measurement of $R$ has an uncertainty of 0.15\%. State lifetimes can also be derived from molecular or photoassociation spectroscopy \cite{Fal06,Wan97}. However, these spectroscopy experiments \cite{Fal06,Wan97} do not distinguish between the $4p_{1/2}$ and $4p_{3/2}$ state lifetimes (they depend on an average) so they cannot be used to determine $R$ nor $\lambda_\textrm{zero}$.

To measure the magic-zero wavelength, we focused 500 mW of laser light asymmetrically on the paths of our three grating Mach-Zehnder atom interferometer \cite{Hol10,Cro09,Ber97}. Atom-waves propagating along each interferometer path acquired a phase shift $\phi(\omega)$ proportional to the dynamic polarizability $\alpha(\omega)$ at the laser frequency $\omega$. We found the laser frequency $\omega_\textrm{zero}=2\pi c/\lambda_\textrm{zero}$ at which the dynamic polarizability vanishes by measuring the phase shift as a function of laser wavelength.

\begin{figure}
\includegraphics{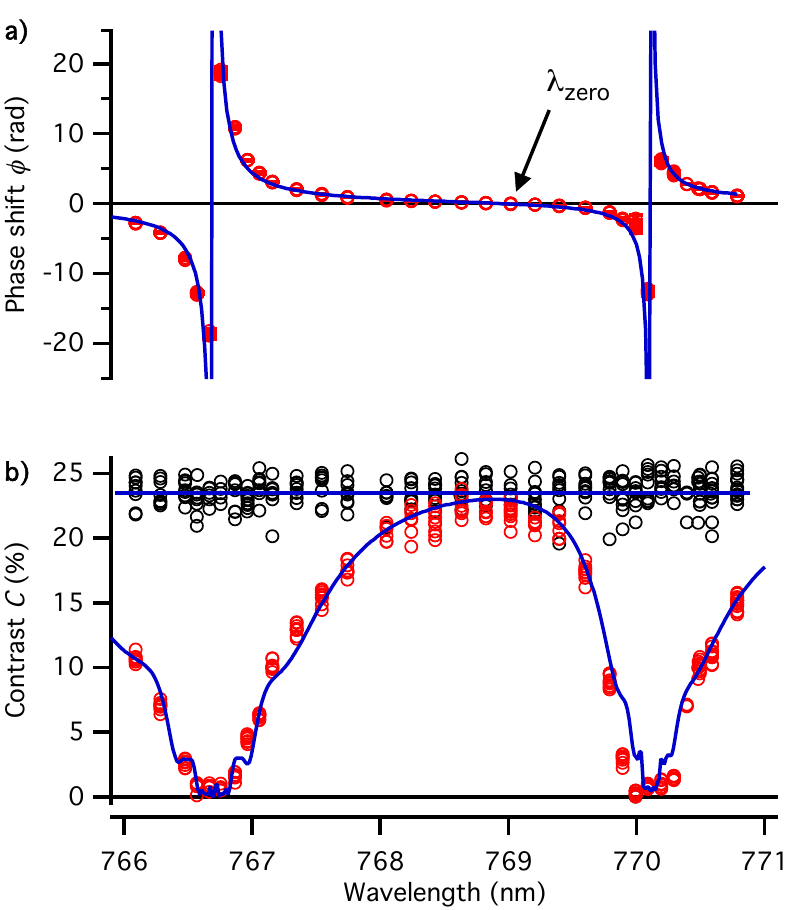}
\caption{\label{TOWfitBig}(Color online) Measurements of the interferometer a) phase shift $\phi$ and b) contrast $C$ as a function of laser wavelength. The measured phase shifts are normalized by the laser power at that wavelength. The reference contrast $C_0$ is shown in black circles.}
\end{figure}

The phase shift $\phi_0(\omega)$ along one interferometer path is given by
\begin{equation}
\label{inteqn}
\phi_0(\omega) = \frac{\alpha(\omega)}{2\epsilon_0 c \hbar v}\int_{-\infty}^{\infty}I(x,z) dz
\end{equation}
where $v\approx1600$ m/s is the atom velocity, $I(x,z)$ is the laser beam intensity (assumed to be monochromatic for now), $x$ is the transverse coordinate in the plane of the interferometer, and $z$ is the longitudinal coordinate. The laser beam intensity was 400 W/cm$^2$ (500 mW focused to a beam waist of $\approx200$ $\mu$m). We measure the differential phase shift $\phi(\omega)$ for two components of the atomic wave functions that are separated by $60$ $\mu$m in our atom interferometer. Figure \ref{TOWfitBig} shows the differential phase shift and contrast of the interferometer as the laser wavelength is scanned 5 nm across the D1 and D2 lines.

\begin{figure}
\includegraphics{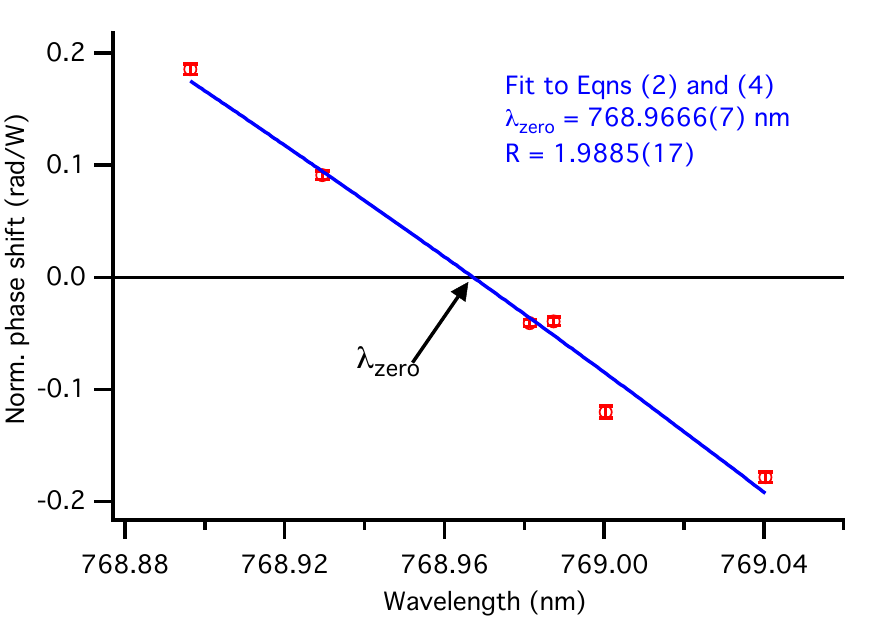}
\caption{\label{TOWfitSmall}(Color online) Measurements of phase shift and laser wavelength. Each point represents 5 minutes of data.  The fit uses Eqs.~(\ref{dynpol}) and (\ref{phibalpha}) described in the text, with free parameters $R$ and $b$.  $R$ determines $\lambda_\textrm{zero}$.}
\end{figure}

Equation (\ref{inteqn}) is useful for understanding the origin of the phase shift, similar to $\phi(\omega)$ shown in \cite{Dei08, Mor93}. But our measurements of $\lambda_\textrm{zero}$ do not depend on precise knowledge of the atom beam velocity nor the focused laser beam irradiance. Changes in these parameters would only affect the magnitude of the phase shift, not the zero crossing. Therefore, we reduce Eq.~(\ref{inteqn}) to simply
\begin{eqnarray}
\label{phibalpha}
\phi(\omega)=b\alpha(\omega),
\end{eqnarray}
where $b$ is a parameter proportional to the laser beam intensity and the interaction time. To precisely measure $\lambda_\textrm{zero}$, we studied phase shifts within 100 pm of $\lambda_\textrm{zero}$, as shown in Figure \ref{TOWfitSmall}. The laser power changed with wavelength and drifted over time, so we monitored the power incident on the atom beam and normalized the measured phase shifts. We reproduced this 1 hr experiment 35 times over a period of 5 days. We fit these data to Eqs.~(\ref{dynpol}) and (\ref{phibalpha}), with $R$ and $b$ as the only free parameters. The precision with which we can determine $\lambda_{\textrm{zero}}$ is inversely proportional to the slope $d\phi/d\lambda$. This slope is typically 1 mrad/pm, and our phase uncertainty from shot noise is $\delta\phi\approx$ 1 mrad with 5 minutes of data.

Our reported measurement of the magic-zero wavelength is the average of 35 individual measurements of $\lambda_{\textrm{zero}}$ similar to the one shown in Figure \ref{TOWfitSmall}, after discarding the highest and lowest 10\% of the measurements. The reported statistical error (0.7 pm) is the standard error of the mean of the trimmed data set. Table \ref{ErrorTable} shows a summary of the error budget and we discuss systematic errors associated with the laser system below.

We generated 2 W of laser light using a MOPA system \cite{Bol10, Nym06}. We used a Littrow ECDL with wavelength-dependent pointing compensation \cite{Haw01} to keep the seed light well-coupled into a tapered amplifier over a 5 nm tuning range. A Bristol Instruments 621B wavelength meter with an accuracy of 0.75 ppm measured the vacuum wavelength of the seed laser.

After spatial filtering with a single mode fiber, 1\% of the power is in a broadband spectral component from spontaneous emission in the tapered amplifier \cite{Voi01}.    To quantify the uncertainty in $\lambda_\textrm{zero}$ caused by this broadband component, we characterized the laser spectrum with a grating spectrometer and we accounted for the laser spectrum by modifying Eq.~(\ref{inteqn}) with an additional integral over the frequency dependent laser intensity.  We calculated that the broadband light introduces an uncertainty of 0.5 pm to our measurement of $\lambda_\textrm{zero}$.

We also measured the crossing angle between the laser and atom beams, and applied a 0.56(5) pm correction to $\lambda_\textrm{zero}$ due to the Doppler shift.  Finally, we calculated that at the intensity we are using, the hyperpolarizability of the ground state causes a shift for $\lambda_\textrm{zero}$ on the order of 0.001 pm. This is negligible in our current experiment but suggests an interesting opportunity for future measurements of intensity-dependent shifts in $\lambda_\textrm{zero}$ due to higher order effects.

\begingroup
\begin{table}
\caption{\label{ErrorTable}Magic-zero wavelength error budget.}
\begin{center}
\begin{tabular}{l c}
\hline\hline
Source & $\lambda_\textrm{zero}$ err. (pm) \\
\hline
Laser wavelength & 0.6\\
Broadband light & 0.5\\
Polarization & 0.1\\
Doppler shift & 0.05\\
\\
Total systematic error & 0.8 \\
Total statistical error & 0.7 \\
\\
Total error & 1.1\\
\hline\hline
\end{tabular}
\end{center}
\end{table}
\endgroup

Contrast loss due to several factors analogous to inhomogeneous broadening limits the precision with which $\lambda_\textrm{zero}$ can be measured. Averaging over the thickness of the atom beam and accounting for +1 and -1 diffraction orders from the 1st nanograting explains most of the observed contrast loss in Figure \ref{TOWfitBig}(b). The velocity spread of the atom beam ($\sigma_v\approx v_0/15$) slightly reduces the observable contrast as well. The small contrast loss due to light at $\lambda_\textrm{zero}$ can be explained by unintended elliptical polarization of the laser beam. Circular polarization causes different Zeeman substates ($m_F$) to acquire different phase shifts even at $\lambda_\textrm{zero}$. Averaging over the 8 $|F,m_F\rangle$ states in our experiment reduces the contrast but introduces little error to $\lambda_\textrm{zero}$ thanks to the equal (thermally distributed) populations of all $m_F$ in our atom beam. We allow for a conservative 0.1 pm uncertainty in $\lambda_\textrm{zero}$ due to unaccounted for effects such as quadratic Zeeman shifts or optical pumping compounded with the light polarization.

Because of the contrast loss from all these mechanisms, if we could optimize our experiment just by increasing the laser power without bound, we would only choose 10 times more power.  Furthermore this would only result in 5 times better sensitivity, approaching 50 $\textrm{pm}/\sqrt{\textrm{Hz}}$.  If we had power to spare, one way to maintain higher contrast would be to use a triangular mask for a large area light beam.  This would cause the differential phase shift to be independent of position in the atom beam.

Next, we explore how photon scattering, analogous to homogeneous broadening, imposes a fundamental limit on the precision with which any magic-zero wavelength can be measured, even in different types of experiments. Atom interferometers are in principle ideal tools for studying the small energy shifts that result from light near $\lambda_\textrm{zero}$.  However, magic-zero wavelengths may also be measured with other methods.  For example, atom loss rates in an optical dipole trap would increase near $\lambda_\textrm{zero}$.  A Bose-Einstein condensate imprinted by a light beam redder (or bluer) than the magic-zero wavelength may produce light (or dark) solitons.  Atoms can diffract from an optical lattice near (but not at) $\lambda_\textrm{zero}$, and atom beam deflections can be induced by light detuned from $\lambda_\textrm{zero}$ \cite{Gou05}.   But all of these methods essentially rely on changes to the center of mass motion for atoms, or equivalently, changes to the de Broglie wave that represents this motion.  Atomic clocks provide similar (picometer) precision for measurements of the magic wavelengths ($\lambda_\textrm{magic}$) that depend on the differential light-shift for two states \cite{Der11,Lud08}, but because clocks are affected by shifts in both ground and excited states, they are less ideal for measurement of magic-zero wavelengths ($\lambda_\textrm{zero}$) discussed here. Furthermore, all of these proposed experiments are limited by decoherence or heating due to photon scattering.

To quantify this fundamental limitation due to decoherence in our experiment, let $\Delta$ be the detuning from resonance, $\Omega$ be the Rabi frequency, and $T$ be the time an atom is exposed to the laser beam. In the large detuning limit ($\Delta^2 \gg \Omega^2$) the slope $d\phi/ d\lambda$ is proportional to $IT/ \Delta^2$ whereas the phase uncertainty increases exponentially with the same factor \footnote{Using the rotating wave approximation $\phi(\omega) \propto IT/\Delta$ and $d\phi/d\omega\propto IT/\Delta^2$. Due to photon scattering, the contrast $C \propto \exp(-IT/\Delta^2)$, and $\delta \phi \propto C^{-1}$.}. This indicates that a more powerful laser or a longer interaction time offers diminishing returns for the experimental sensitivity to $\lambda_\textrm{zero}$. To minimize the shot noise limited uncertainty in $\lambda_{\textrm{zero}}$ we should increase the pulse area ($IT$) until we obtain a contrast reduction of $C/C_0 = e^{-1}$. 

Our experiment could be significantly improved by increasing the atom interferometer path separation so the laser can be entirely focused (with homogeneous irradiance) on one interferometer path.  In this more ideal situation,  decoherence is the only remaining source of contrast loss.  Then the maximum achievable slope $d\phi/d\lambda$ is found via
\begin{eqnarray}
\label{TOWlimit}
\frac{d\phi}{d\omega}\approx\frac{1}{2\Gamma}P_\textrm{s}
\end{eqnarray}
where $P_\textrm{s}$ is the probability that atom scatters one or more photons and $\Gamma$ is the excited state decay rate.  With optimized contrast loss due to scattering ($P_\textrm{s} = 1-e^{-1}$) the slope becomes as large as $d \phi / d \lambda = $ 40 rad/pm.   In this way, future measurements of magic-zero wavelengths can be made with very high precision, possibly with accuracy limited by a shot noise sensitivity better than picometers per $\sqrt{\textrm{Hz}}$ with current technology.  Perhaps this can be achieved in an ultracold atom interferometer \cite{Dei08}, however such experiments typically would measure the magic-zero wavelength of a particular $|F,m_F\rangle$ state and therefore may be more sensitive to uncertainties in the laser polarization and magnetic fields.

As an outlook, the $\lambda_\textrm{zero}$ measurement presented here provides a foundation for a new set of experimental benchmarks that can be used to test atomic structure calculations. Future measurements of several other magic-zero wavelengths in potassium and other atoms can be accomplished with similar techniques. For example, in potassium atoms, two additional magic-zero wavelengths occur near 405 nm. These $\lambda_\textrm{zero}$ are primarily determined by transitions from $4s$ to $4p_{1/2}$, $4p_{3/2}$, $5p_{1/2}$, and $5p_{3/2}$ states. Measurements of two other $\lambda_\textrm{zero}$ combined with the one reported here could therefore be used to specify ratios of four line strengths.  However, $\alpha_\textrm{core}$ (the largest component of the semi-empirical parameter $A$ in Eq.~(\ref{dynpol})) is no longer negligible for $\lambda_\textrm{zero}$ near 405 nm \cite{Saf12per}. Therefore, new $\lambda_\textrm{zero}$ measurements will also provide benchmark tests for the contributions from core electrons to polarizabilities. Magic-zero wavelength measurements in heavier atoms, where the fine-structure splitting is larger, will be more sensitive to both core-electron contributions and relativistic corrections to the line strength ratio $R$. Measurements of hyperpolarizability may also be accomplished by measuring energy shifts at magic-zero wavelengths that depend on intensity-squared (i.e. $E^4$).

In summary, we measured the longest magic-zero wavelength of potassium with 1 pm uncertainty. The measured phase shifts and resulting precision in $\lambda_\textrm{zero}$ could be increased by 3 orders of magnitude in future work by focusing a laser beam entirely on one path of the atom interferometer, more accurate measurements of the laser spectrum, and more careful control of the laser polarization.

We thank M.S. Safronova for enlightening discussions. This work is supported by NSF Grant No.~0969348 and a NIST PMG. WFH thanks the Arizona TRIF and RT thanks NSF GRFP Award No.~DGE-1143953 for additional support.

\bibliography{../../../MasterBibliography8}
\end{document}